  \documentclass{emulateapj}
\newcommand{\philtrans} {Phil.\ Trans.\ Roy.\ Soc.\ Lond.}
\newcommand{\naturwiss} {Naturwiss.}
\newcommand{\za}        {Zeitschrift f.\ Astrophys.}

\newcommand{\DEM}{{\rm{DEM}}}
\newcommand{\EM}{{\rm{EM}}}
\newcommand{\eqn}[1]{(\ref{#1})}
\newcommand{\fig}[1]{Fig.\,\ref{#1}}
\newcommand{\tab}[1]{Table\,\ref{#1}}
\newcommand{\sect}[1]{Sect.\,\ref{#1}}
\newcommand{\bm}[1]{\mbox{\boldmath$#1$\unboldmath}}
  
  \def\figwtxt{\textwidth} \def\figwcol{\columnwidth}

\journalinfo{{\rm \today \hspace*{\fill} Submitted to~ {\sc The Astrophysical Journal}~~}}

\shorttitle{Forward modeling of stellar coronae}
\shortauthors{Peter, Gudiksen \& Nordlund}

\begin{document}


\title{Forward modeling of stellar coronae:\\
       from a 3D MHD model to synthetic EUV spectra}

\author{Hardi Peter}
\affil{Kiepenheuer-Institut f\"ur Sonnenphysik, 79104 Freiburg, Germany;
       {peter@kis.uni-freiburg.de}}

\author{Boris V. Gudiksen}
\affil{Institute of Theoretical Astrophysics, University of Oslo, Norway}

\author{{\AA}ke Nordlund}
\affil{Niels Bohr Institute, University of Copenhagen, Denmark}

\begin{abstract}
%
A forward model is described in which we synthesize spectra from an
\emph{ab-initio} 3D MHD simulation of an outer stellar atmosphere, where
the coronal heating is based on braiding of magnetic flux due to
photospheric footpoint motions.
We discuss the validity of assumptions such as ionization equilibrium
and investigate the applicability of diagnostics like the differential
emission measure inversion.
We find that the general appearance of the synthesized corona is similar
to the solar corona and that, on a statistical basis, integral quantities
such as average Doppler shifts or differential emission measures are
reproduced remarkably well.
The persistent redshifts in the transition region, which have puzzled
theorists since their discovery, are explained by this model as caused
by the flows induced by the heating through braiding of magnetic flux.
While the model corona is only slowly evolving in intensity, as is
observed, the amount of structure and variability in Doppler shift is very large.
This emphasizes the need for fast coronal spectroscopy, as the dynamical
response of the corona to the heating process manifests itself in
a comparably slow evolving coronal intensity but rapid changes in
Doppler shift.%
\footnote{{%
Movies showing the temporal evolution of the synthesized appearence of the
model corona similar to \fig{F:image} and \fig{F:cut} can be found on the web-site
http//www.kis.uni-freiburg.de/$\sim$peter/movie/corona\_spec/.
}}
\end{abstract}
\keywords{      MHD
            --- stars: coronae
            --- Sun: corona
            --- Sun: UV radiation}

\section{Introduction}                                  \label{S:intro}

Since the discovery that the outer atmosphere of the Sun, the corona, is
much hotter than the photosphere \citep{Grotrian:1939,Edlen:1942} it
remains a mystery what heats the coronae of the Sun and other cool stars
to more than a million K.
During the last decade a wealth of coronal data was collected through
remote sensing, especially with the help of the Yohkoh, SOHO and TRACE
space missions.
Since the beginning of coronal physics there have been countless
suggestions of coronal heating processes, as to be found in the
proceedings of conferences on coronal heating, e.g.\
\cite{Ulmschneider+al:1991} or the SOHO 15 meeting \citep{Walsh_soho:2004}.
There was never a lack of suggestions --- the problem is how to prove
or disprove a model with the help of observations.

Observations of the corona provide us with the flux and energy of photons.
To test a model one might use an inversion of the observations to derive
e.g.\ temperatures and densities, and compare this to the modeled plasma
parameters.
In such an inversion procedure one has to make implicit assumptions and
often the inversion problem is ill-posed
\citep[e.g.][]{Judge+McIntosh:1999,McIntosh:2000}.
The other major approach is forward modeling.
Based on a model one derives not only the plasma parameters, but also the
photon spectra resulting from the model atmosphere.
Like an inversion, also a forward model is based on assumptions, but
usually the assumptions are more easy to control in a forward model
approach.
In the past numerous forward models have been applied to study various
coronal phenomena, like transition region redshifts
\citep[e.g.][]{Hansteen:1993}, the connectivity of the atmosphere
\citep[e.g.][]{Wikstol+al:1998}, or catastrophic
cooling in loops \citep[e.g.][]{Mueller+al:2004a}.
Likewise the inversion approach can allow a detailed comparison, when
treated with care.
For example \cite{Priest+al:1998} compared the temperature profiles from
various loop heating models and compared these to the temperature profile
derived for a large isolated loop as inverted from Yohkoh soft X-ray
observations.

There are many observational challenges a coronal model has to explain,
and we will here only name a few which are of relevance for our work.
The structures are much more smooth in the corona than in the transition
region \citep[e.g.][]{Reeves:1976}.
The emission measure, i.e. the intrinsic capability for a volume at a
given temperature to radiate, is strongly increasing from the transition
region down to the chromosphere \citep[e.g.][]{Dowdy+al:1986}.
The middle transition region shows persistent redshifts
\citep{Doschek+al:1976}, while the low corona shows a net blueshift
\citep{Peter:1999full}.
The unresolved motions are largest in the middle transition region
\citep[e.g.][]{Chae+al:1998:width}.
The temporal variability of the line intensities is largest in the middle
transition region \citep[e.g.][]{Brkovic+al:2003}.
Here we cannot give a full list of references relevant for coronal
heating based on data from Yohkoh, SOHO and TRACE,
but we would like to emphasize the usefulness of spectroscopic
investigations using the SUMER EUV spectrometer on SOHO
\citep{Wilhelm+al:1995,Wilhelm+al:1997}

In the light of the new observations provided in the recent years it is
especially encouraging that complex three-dimensional \emph{ab-initio}
models for the corona based on magnetohydrodynamics (MHD) are becoming available
\citep{Gudiksen+Nordlund:2002,Gudiksen+Nordlund:2005a,Gudiksen+Nordlund:2005b}.
In this work we calculate the EUV spectra from these models, which allows
us to perform a detailed comparison to observations.
The aim of this paper is to describe the validity and applicability of the
approach chosen here to synthesize the EUV spectra and to show the
huge potential of this forward modeling approach.
First results on the spectral synthesis have already been published in
\cite{Peter+al:2004}.

We will first give a very quick introduction to the coronal MHD model
(\sect{S:MHD}), before we describe the synthesis of the EUV spectra in
\sect{S:synthetic}.
The validity and applicability of our approach will be discussed in
\sect{S:Validation} and finally in \sect{S:analysis} we will investigate
the relation of our approach to real solar observations, like the
morphology, average line shifts and widths or the differential emission
measure.

\section{3D MHD model of coronal structures}            \label{S:MHD}

Here we will only give a very brief overview on the basics of the 3D MHD
model underlying our spectral synthesis.
A detailed description of the model can be found in
\citet{Gudiksen+Nordlund:2005a,Gudiksen+Nordlund:2005b}.

In this model the heating of the corona is due to braiding of magnetic
flux as a consequence of photospheric motions, as suggested first by
\cite{Parker:1972}.
The braided magnetic field gives rise to currents which are then
dissipated through Ohmic heating.
The horizontal motions in the photosphere are constructed based on a
Voronoi-tessellation \citep[e.g.][]{Okabe+al:1992}.
This flow field reproduces the geometrical pattern and the amplitude power
spectrum of the velocity and the vorticity.

The computational box contains a volume of $60\times 60$\,Mm$^2$ in the
horizontal directions and 37\,Mm vertically covering the whole atmosphere
from the photosphere to the corona with a non-equidistant grid of 150$^3$
points.
A 5'th order in space, 3rd order in time,
fully compressible 3D magneto-hydro-dynamics (MHD) code on a
staggered non-equidistant mesh is used, including classical heat conduction
along the magnetic field \citep{Spitzer:1956} and a cooling function of
optically thin radiative losses.
A Newtonian cooling scheme is used to keep the atmosphere near a
prescribed temperature profile in the photosphere and
chromosphere.

The simulation starts with an initial condition with a magnetic field from
a potential field extrapolation based on an observed magnetogram, scaled
down to fit into the computational box.
The heating through braiding of magnetic flux rapidly leads to a corona
which is intermittent in space and time, typically reaching temperatures
of about $10^6$\,K.
On average the heating is concentrated very much at the bottom of the
computational domain in the chromosphere and drops exponentially towards
larger heights.
The heat flux into the corona of about 2000--8000\,W/m$^2$ is comparable to
the energy losses derived from observations
\citep[e.g.][]{Withbroe+Noyes:1977}.
The amount of heating in this model is only a lower limit, as the heating
would stay constant or increase with increased spatial resolution
\citep{Hendrix+al:1996,Galsgaard+Nordlund:1996}, and the initial conditions
and the driver are constructed in a way as to induce a minimum of stress in
the magnetic field \citep[see discussion in][]{Gudiksen+Nordlund:2005a}.

For the spectroscopic investigations in this paper we use only the last
${\sim}20$\,minutes of the whole simulated time span of
${\sim}50$\,minutes, to ensure that our results are sufficiently independent
of the initial conditions.
Thus in this paper the time $t{=}0$\,minutes refers to a time
${\sim}30$\,minutes into the MHD simulation.

\section{Calculation and analysis of synthetic spectra}
                                                        \label{S:synthetic}

To study the spectral line properties from the MHD calculation we
first calculated the emissivities of a number of emission lines from the
transition region and low corona at each grid point.
Using these emissivities and the velocities from the MHD calculation we
calculate the spectrum for each line at each grid point.
Finally we integrate the spectra along a line-of-sight in order to
obtain spatial maps of line intensity, shift and width.
Similarly we also compute the average spectra from the box to study the
average emission line properties.
Thus the spectra synthesized from the MHD calculation are directly
comparable to observations of EUV spectra, on a statistical basis, of
course.

The lines used in this paper are listed in \tab{T:lines}.
All these lines are optically thin, at least at disk center ---
only this allows the concept of synthesizing the spectra as outlined
above without accounting for radiative transport (see also
\sect{S:Validation}).
The lines have been chosen using two criteria.
Firstly we wanted to span the whole range of temperatures in the transition
region from the chromosphere to the corona, i.e. from 10$^4$ to 10$^6$~K.
And secondly the lines should be observable with SUMER to allow for a
comparison with the observed properties of the line profile.

\subsection{Line emissivity}                    \label{S:line_emission}

The line emissivities at each grid point are calculated under the
assumption of ionization equilibrium (cf. \sect{S:ionisation}).
To calculate the various populations of excitation and ionization states we
use the atomic data package CHIANTI \citep[Version 4.02;
see][]{Dere+al:1997,Young+al:2003}.
Because of computational time we first create a look-up table for a large
number of temperatures and densities, and then read from this table to
extract the line emissivities.

The lines considered in this paper (\tab{T:lines}) are
predominantly excited by electron collisions and de-excited by spontaneous
emission.
The emissivity (integrated over the line profile, i.e.\ energy per time
and volume) is then given by
\begin{equation}\label{E:emissivity}
\varepsilon = h\nu \, n_2 A_{21} ~,
\end{equation}
where $h\nu$ is the energy of the transition, $n_2$ is the population
number density of the upper level and $A_{21}$ is the Einstein coefficient
for spontaneous emission to the lower level.
Eq.\ \eqn{E:emissivity} can be rewritten as
\begin{equation}\label{E:emissivity_rel}
\varepsilon = G(T,n_e)~n_e^2 ~ ,
\end{equation}
with the electron number density $n_e$ and a function $G(T,n_e)$ defined as
\begin{equation}\label{E:goft}
G(T,n_e) = h\nu\ A_{21}
            \cdot  \frac{n_2}{n_e~n_{\rm{ion}}}
            \cdot  \frac{n_{\rm{ion}}}{n_{\rm{el}}}
            \cdot  \frac{n_{\rm{el}}}{n_{\rm{H}}}
            \cdot  \frac{n_{\rm{H}}}{n_e}
            ~ .
\end{equation}
The terms describe the excitation of the line, the ionization of the
respective ion, the elemental abundance and the degree of ionization of
the plasma.
The number density of ions in the respective ionization state is
$n_{\rm{ion}}$, the number density of the respective element is
$n_{\rm{el}}$ and $n_{\rm{H}}$ denotes the number density of hydrogen.

The first term of \eqn{E:goft}, $h\nu{}A_{21}$, is a constant given by atomic physics.

Because the excitation of the lines considered here is due to electron
collisions, the population of the upper level $n_2/n_{\rm{ion}}$ is basically
proportional to the electron density.
This is true only for density insensitive lines and for a certain range of
densities.
Thus the second term of \eqn{E:goft} depends mainly on temperature, but
generally also on density.
We fully account for the temperature and density dependence when
calculating this term using CHIANTI.

Under the assumption of ionization equilibrium the ionization degree
$n_{\rm{ion}}/n_{\rm{el}}$ in \eqn{E:goft}
depends only on temperature and we use the values calculated by
\citet{Mazzotta+al:1998} as tabulated in the CHIANTI package.
Of course, this is a strong assumption in a dynamic atmosphere like
investigated in the present MHD model.
Nevertheless in large parts of the atmosphere under investigation here
this assumption can be considered a good first step.
This will be discussed separately in \sect{S:ionisation}.

In evaluating Eq.\,\eqn{E:goft} we use constant abundances
$n_{\rm{el}}/n_{\rm{H}}$ throughout the computational box and adopt the most
recent solar photospheric values as tabulated in the CHIANTI package.
Especially we do not account for the change of abundances (by a factor of
2 to 3) that is to be expected in the chromosphere, i.e.\ the FIP effect
\citep[e.g.][]{Steiger+al:2000}.

In a fully ionized plasma with hydrogen and 10\% Helium the value of
$n_{\rm{H}}/n_e$ is about 0.8. Here we use CHIANTI to calculate
this term depending on the degree of ionization, i.e. on temperature.

From this discussion we see that the contribution function $G(T,n_e)$ as
defined in \eqn{E:goft} depends only weakly on density when selecting an
appropriate line.
Thus it is often referred to as $G(T)$.%
\footnote{%
Please note that $G(T,n_e)$ is defined differently by different authors;
e.g.\ often the abundance is not included in the definition of $G$.
}
Mainly because of the ionization equilibrium the function $G(T,n_e)$
strongly peaks in temperature.
Based on the ionization equilibrium calculations a canonical value for the
width of the contribution function is 0.3 in $\log{}T$
(cf.\ \sect{S:lineformation}).
For density insensitive lines, i.e. $G{=}G(T)$, the emissivity
\eqn{E:emissivity_rel} reflects that the optically thin radiative losses
are proportional to the density squared.

To evaluate the emissivity \eqn{E:emissivity_rel} at each grid point of the
MHD calculation we finally compute the electron density $n_e$ from the mass
density $\rho$ of the MHD calculation using the degree of ionization, again
as obtained from CHIANTI.
By an integration along the line-of-sight we can then calculate the energy
flux density in a given emission line out of our computational box, e.g.\
in units of W/m$^2$
These can be easily converted into line radiances as observed e.g.\ at the
location of SOHO.

\begin{table}[t]
\begin{center}
\caption{%
Emission lines synthesized in this study.
\label{T:lines}}
\begin{tabular}{l@{~}rc@{~~}c@{~~}ccc}
\hline
\hline
        &
        & \multicolumn{4}{c}{formation temperature: log~$T$\,[K]}
        & FWHM
\\
\cline{3-6}
        &
        & ion.eq.
        & excitation
        & \DEM
        & MHD
        & $\Delta \log T$
\\
\hline
\ion{Si}{2}  & 1533 & 4.60 & 4.36 & 4.08 & 4.25 & --
\\
\ion{Si}{4}  & 1394 & 4.81 & 4.85 & 4.82 & 4.90 & 0.29
\\
\ion{C}{2}   & 1335 & 4.67 & 4.57 & 4.23 & 4.64 & 0.22
\\
\ion{C}{3}   &  977 & 4.78 &  --  & 4.71 & 4.84 & 0.29
\\
\ion{C}{4}   & 1548 & 5.00 & 5.00 & 5.02 & 5.11 & 0.25
\\
\ion{O}{4}   & 1401 & 5.27 & 5.14 & 5.15 & 5.18 & 0.32
\\
\ion{O}{5}   &  630 & 5.38 & 5.35 & 5.40 & 5.44 & 0.28
\\
\ion{O}{6}   & 1032 & 5.45 & 5.44 & 5.50 & 5.60 & 0.23
\\
\ion{Ne}{8}  &  770 & 5.81 & 5.76 & 5.82 & 5.89 & 0.16
\\
\ion{Mg}{10} &  625 & 6.04 & 6.01 & 6.01 & 6.06 & 0.17
\\
\hline
\end{tabular}
\end{center}
{
\footnotesize
The line formation temperatures given here are as following from different
methods: maximum of ionization fraction (``ion.eq.''), maximum of
ionization fraction accounting for collisional excitation (``excitation''),
constant pressure {\DEM} inversion using CHIANTI (``DEM'') and based on
emissivities of the present MHD model (``MHD'').
The rightmost column gives the width of the respective contribution
function in temperature as following from the present work in a
logarithmic scale.
See \sect{S:lineformation} for a detailed discussion.
}
\end{table}

\subsection{Synthetic spectral profiles}                \label{S:spectrum}

For convenience we calculate all spectral profiles with the wavelength
given in units of (Doppler) velocities.

To calculate the emission line profiles at each grid point we assume that
the line profile has a thermal width $w_{\rm{th}}$ according to the
temperature $T$ as obtained from the MHD calculation at the respective
grid point,
\begin{equation}\label{E:thermal_width}
w_{\rm{th}} = \left( \frac{2\,k_{\rm{B}}~T}{m_i} \right)^{1/2} ~.
\end{equation}
Here $k_{\rm{B}}$ is Boltzmann's constant and $m_i$ is the atomic mass of
the respective ion.

Furthermore the line is Doppler shifted by the line-of-sight velocity at
the respective grid point.
In this paper we concentrate on a vertical line-of-sight, as we would like
to compare our results of average synthesized Doppler shifts to
observations of solar disk center Doppler shifts.
Thus we use the vertical component of the velocity $v_{\rm{v}}$ from the
MHD calculation.

Now the line profile $I_v$ at each grid point is given by
\begin{equation}\label{E:line_profile}
I_v = I_{\rm{peak}} \exp \left( -\,\frac{(v-v_{\rm{v}})^2}{w_{\rm{th}}^2} \right) ~.
\end{equation}
For such a Gaussian with an exponential width $w_{\rm{th}}$ the peak
intensity $I_{\rm{peak}}$ is related to the total intensity
$I_{\rm{tot}}$, i.e.\ integrated over wavelength (or here velocity), by
\begin{equation}\label{E:total_peak}
I_{\rm{tot}} = \sqrt{\pi} \, I_{\rm{peak}} \, w_{\rm{th}} ~ .
\end{equation}
The total intensity $I_{\rm{tot}}$ is given through the emissivity
\eqn{E:emissivity_rel} as discussed in \sect{S:line_emission}.

Thus we can use \eqn{E:line_profile} to calculate the spectrum at each grid
point as viewed along the (vertical) line-of-sight.
To obtain the total spectrum we integrate along the line-of-sight,
\begin{equation}\label{E:trapez}
I_v^{\rm{synth}} = \int_{\rm{line-of-sight}} I_v ~ {\rm{d}}l ~.
\end{equation}
This yields two dimensional maps of spectra as they would be obtained if
an EUV spectrometer like SUMER would perform a raster scan.
Thus we name these spectra $I_v^{\rm{synth}}$, {\em synthetic spectra}.
These are (in general) not Gaussian, even though in the present study they
are mostly close to Gaussian.


The analysis of the synthetic spectra is similar to the procedure for
observed spectra: we calculate line intensity, position and width of
$I_v^{\rm{synth}}$ at each location of the spatial maps.
For observed solar spectra it is preferable (and almost always necessary) to
perform a Gaussian fit to derive these parameters, which is mainly because
of the noise in the data.
As our synthetic spectra are noise-free and very close to Gaussians it is
sufficient to calculate the profile moments with respect to wavelength (or
here velocity $v$).
Then line intensity, shift and width are given by
\begin{eqnarray}
I^{\rm{synth}} &=& \int I_v^{\rm{synth}} ~{\rm{d}}v ~,
\\
v^{\rm{synth}} &=& \frac{1}{I^{\rm{synth}}} \int v~I_v^{\rm{synth}} ~{\rm{d}}v
   ~,
\\
w^{\rm{synth}} &=& \frac{1}{I^{\rm{synth}}}
                 \left(2\int  (v-v^{\rm{synth}})^2
                              ~ I_v^{\rm{synth}} ~{\rm{d}}v \right)^{1/2} ~.
\end{eqnarray}
It is easy to check that this indeed returns the desired parameters if
$I_v^{\rm{synth}}$ whould be exactly Gaussian.
Examples of maps of synthesized Doppler shifts and intensities are shown
in \fig{F:image} (panels a/$\alpha$ and b/$\beta$,
respectively; panels c/$\gamma$ and d/$\delta$ show the box as seen from the
sides in intensity; Latin and Greek letters label maps about 7 minutes
apart in time).

When analyzing the line width in observations one usually subtracts the
thermal line with as defined in \eqn{E:thermal_width} assuming the
line is formed at a single temperature $T_{\rm{line}}$.
Here we use the temperatures as derived from the analysis of line formation
temperatures (see \sect{S:lineformation}; column ``MHD'' in \tab{T:lines}).
This results in the non-thermal line width or non-thermal velocity
$w^{\rm{synth}}_{\rm{nt}}$ characterizing the unresolved motions.
This includes the velocity fluctuation along the line-of-sight.

\begin{equation}\label{E:nonthermal}
w^{\rm{synth}}_{\rm{nt}} = \left( \left(w^{\rm{synth}}\right)^2 ~-~
                           \frac{2\,k_{\rm{B}}~T_{\rm{line}}}{m_i}
                                \right)^{1/2}
        ~.
\end{equation}
It is important to note that when synthesizing the spectrum at a given
grid point we use the temperature at that grid point to evaluate the
thermal width following \eqn{E:thermal_width}.
After integration along the line-of-sight we subtract the thermal width in
\eqn{E:nonthermal} as to be expected at an average $T_{\rm{line}}$ in the
line formation region.
Thus our analysis of the data follows closely the analysis of ``real''
observational data.

\subsection{Differential emission measure analysis}             \label{S:dem}

The concept of the (differential) emission measure ({\DEM} or {\EM}) is widely
used in the analysis of stellar and solar optically thin plasmas.
Thus we will apply this tool also to our synthetic spectra.

Following the discussion in \sect{S:line_emission} the line flux (i.e.\
the energy flux of the photons, e.g.\ in W/m$^2$) at a given location on
the Sun is given by a height  integration of the emissivity
\eqn{E:emissivity_rel}.
\begin{equation}\label{E:total_flux}
F ~=~ \int G(T,n_e) ~ n_e^2 ~ {\rm{d}}h ~ .
\end{equation}
To perform the {\DEM} analysis one has to make the assumption that
the integration over height can be substituted by an integration over
temperature, i.e.
\begin{equation}\label{E:total_flux_dem}
F ~=~ \int G(T,n_e) ~ {\DEM} ~ {\rm{d}}T ~ .
\end{equation}
with the differential emission measure
\begin{equation}\label{E:dem}
{\DEM} ~=~ n_e^2 ~ \frac{{\rm{d}}h}{{\rm{d}}T} ~ .
\end{equation}
It is important to note that this implicitly assumes that the temperature
varies monotonically with height.
The highly structured nature of the corona, of course, proves this
assumption untenable (see \sect{S:dem_mhd} and \sect{S:dem_results}).

Using a suitable set of lines covering a large range of temperatures one
can derive the {\DEM} from the observed or synthesized line intensities.
To account for the density dependence of $G$ one has to
make an assumption on the density in the atmosphere.
We use the approximation of constant pressure, which is most common for
the transition region because of its small thickness compared to the
pressure scale height.
To perform the {\DEM} inversion we employed the CHIANTI package.
This also returns the respective line formation temperatures (see
\sect{S:lineformation}).

We would like to stress here that the inversion of ${\DEM}(T)$ is an
ill-posed problem \citep[e.g.][]{Judge+McIntosh:1999,McIntosh:2000}
and that one has to be very careful in the selection of lines, e.g.\ with
respect to the iso-electric sequence \citep{DelZanna+al:2002}.
However, our emphasis in this paper is not on a best possible ${\DEM}$
inversion --- we are applying the ${\DEM}$ inversion to our data only to show
that our synthetic spectra yield a ${\DEM}$ distribution with temperature
similar to observed ones when applying standard techniques.

\section{Validation of the approach}                    \label{S:Validation}

The spectral synthesis as outlined above presents us with two major
limitations.
The formation of lines at low temperatures is not well representsed and
the ionization equilibrium is a strong assumption.

The MHD model is not including a self-consistent treatment of the
chromosphere including radiative transport, but is using a Newton cooling
mechanism to keep the lower part of the computational box at a prescribed
chromospheric temperature profile.
Thus any diagnostics of the plasma at temperatures below, say, $10^4$\,K
is meaningless, and will not be touched upon in this paper.

For the lines formed in the low transition region, below
$\log{T}{\approx}4.5$, we will be confronted with another problem.
Because of the very steep rise of density from the transition region down
into the chromosphere, through \eqn{E:emissivity_rel} the formation of these lines is shifted towards
considerably lower temperatures, as will be discussed in
\sect{S:lineformation}.
Whether this is an artefact of the MHD model used here, or a general
property of outer stellar atmospheres can only be clarified when more
complex coronal models become available.
If indeed these low temperature lines are formed well below the
temperature of peak ion fraction (cf. \fig{F:lineformation}a), this would
have serious impact on the interpretation of observations of these lines.

For the present study it seems that ionization equilibrium is a good
assumption as outlined in \sect{S:ionisation}.
Despite velocities of some 10 or 20\,km/s in the middle transition region
this assumption holds.
Previous studies found contrary results, but they neglected the
back-reaction of the flow on the temperature gradient.
Future studies with more violent flows, however, will have to consider
departures from ionization equilibrium, as is known e.g.\ from the
modeling of explosive events or reconnection
\citep[e.g.][]{Roussev+al:2001c,Roussev+Galsgaard:2002b}.
Those models, however, where only 2D-MHD models and the non-equilibrium
ionization was solved properly only in 1D along a line-of-sight aligned
with the outflow from a reconnection site.
1D codes with non-equilibrium ionization exist since quite a while
\citep[e.g.][]{Hansteen:1993}, but to move on to a full 3D treatment is a
major step.

\subsection{Line formation temperatures}                \label{S:lineformation}

The function $G$ as used in \eqn{E:emissivity_rel} is a
function sharply peaked in temperature.
Thus one can define a line formation temperature.
As $G(T)$ is mainly determined by the fractional density of the ionization
stage, one often uses the peak temperature of the ionization fraction
$n_{\rm{ion}}/n_{\rm{el}}$.
In \tab{T:lines} values as following from the ionization equilibrium
calculation of \citet{Landini+Fossi:1990} are listed (column ``ion.eq.'').
One can go one step further and also include the temperature dependence of
the collisional excitation process by using the peak of the upper level
population in a simple density independent way
$n_{\rm{ion}}/n_{\rm{el}}\,{\cdot}\,T^{-1/2}\exp(-h\nu/k\,T)$ --- values
obtained by \citet{Chae+al:1998:width} and \citet{Chae+al:1998} are listed
in the column ``excitation'' of \tab{T:lines}.
However, $G$ depends also on density, which is not accounted for by these two
approaches.
A {\DEM} inversion as discussed in \sect{S:dem} returns the line formation
temperatures for a constant pressure atmosphere, implicitly assuming a
simple 1D stratified atmosphere (column ``DEM'' in \tab{T:lines}).

For the analysis of the synthetic spectra from our MHD model we do not have
to make all these simplifying assumptions, but we can calculate directly
the contribution function of the respective lines.
Employing the procedure outlined in \sect{S:line_emission} the emissivity
is calculated at each grid point.
Based hereupon we integrate the total emission from the computational box
in each line in small temperature intervals (actually, we use intervals in
$\log_{10}$ of temperature).
As the atmosphere in our model is highly structured, the
volume each temperature interval covers is not necessarily spatially
connected.
From this we can compute the contribution of the respective line as a
function of temperature.

\begin{figure}[t]
\includegraphics[width=\figwcol]{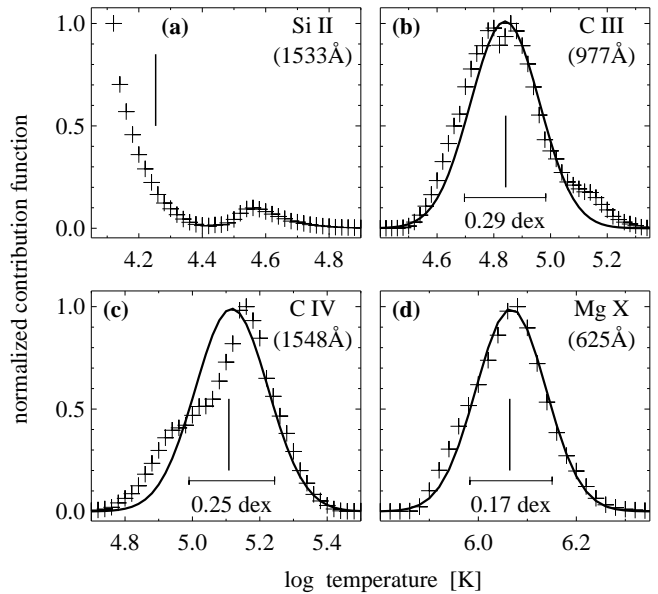}
\caption{%
Contribution to the emissivity as a function of temperature for a number of
lines (crosses).
Except for \ion{Si}{2} the contribution in $\log T$ is roughly
represented by a Gaussian.
Gaussian fits are over-plotted as solid lines and the respective full
widths at half maximum (FWHM) of the Gaussians are given as bars.
The FWHM values are also given on a $\log_{10}$ scale with the bars.
The vertical lines represent the first moment of the contribution
function, used as a line formation temperature in this paper.
See \sect{S:lineformation}.
\label{F:lineformation}}
\end{figure}

These contribution functions are plotted as crosses in
\fig{F:lineformation} for a number of lines.
We have used intervals of 0.02 in $\log_{10}T$, but the results change
only slightly for other values.
Except for \ion{Si}{2} the contribution in $\log{T}$ is roughly
represented by a Gaussian.
To account also for the asymmetries in the contribution function, which are
due to the highly structured nature of the atmosphere, we use the first
moment of the contribution function in $\log{T}$ to describe the line
formation temperature (this is practically identical to the center of the
respective Gaussian fit, cf.\ \fig{F:lineformation}).

A canonical value for the range of temperatures contributing to
transition region lines is about 0.3 in $\log{T}$, i.e.\ 0.3\,dex.
The full width at half maximum (FWHM) of the Gaussian fits is usually a
bit smaller than that, but 0.3\,dex seems to be a good choice, even for a
highly structured atmosphere as considered here (cf.\
\fig{F:lineformation} and rightmost column of \tab{T:lines}).

The contribution function for the ``coolest ion'' discussed in this paper,
\ion{Si}{2}, is far from being Gaussian or even having a well pronounced
peak in temperature (\fig{F:lineformation}a).
There is a small secondary peak just below $\log{T}{=}4.6$, where the
ion fraction peaks (cf.\ \tab{T:lines}), but below $\log{T}{=}4.4$ the
contribution of \ion{Si}{2} rises and well exceeds the value at the
temperature of peak ion fraction.
This is due to the density dependence of the contribution function $G$
(also the {\DEM} inversion using CHIANTI gives a low temperature of
$\log{T}{=}4.08$ for this ion, cf.\ \tab{T:lines}).
The bulk part of the \ion{Si}{2} emission originates from temperatures
below $\log{T}{=}4.4$, where the ion fraction of Si$^+$ is still low.
As there is no self-consistent treatment of the upper chromosphere,
i.e. for temperatures up to some 20\,000\,K ($\log{T}{=}4.3$), the
diagnostic potential of \ion{Si}{2} is limited within the framework of
this study.

The main point of this discussion on \ion{Si}{2} is to show that based on
the present model, it seems
not very useful at all to use \ion{Si}{2} or other comparably cool ions
to extract information on the state of the transition region.
The temperature regime below 20\,000\,K, where also Ly-$\alpha$ is formed,
is not well described by a simple stratified optically thin atmosphere as
assumed for (almost all) observational investigations of EUV emission lines.

\subsection{Ionization balance}                 \label{S:ionisation}

To properly account for ionization effects one would have to solve the
rate equations for each species, i.e.\ for a considerable number of
ionization and excitation states.
In general for an ionization/excitation state $i$ this is
\begin{equation}\label{E:rate_equations}
  \frac{{\rm{d}}n_i}{{\rm{d}}t}
+ \nabla \cdot (n_i\,\bm{v})  =
    \sum_j \left( -\gamma_{ij} + \gamma_{ji} \right) ~,
\end{equation}
where $n_i$ is the number density and $ \bm{v}$ the bulk
velocity.
The summation on the right-hand-side is over all loss
processes to states $j$ and gain processes from $j$ with rates
$\gamma_{ij}$ and $\gamma_{ji}$.

Currently we cannot solve these rate equations parallel to the 3D MHD
problem, simply because of computational time.

Therefore, when calculating the emissivity of the various lines as outlined in
\sect{S:line_emission}, we assume ionization equilibrium.
To check this severe simplification we compare the ionization times to
dynamic times given by the temperature gradient and the velocity of the
plasma.

\subsubsection{Ionization times}                \label{S:ionisation_times}

To calculate the ionization times $\tau_{\rm{ion}}$ at each grid point in
the box of the MHD calculation we use the ionization rates $C_{\rm{ion}}(T)$
as parameterized in \citet{Arnaud+Rothenflug:1985}.
These rates are still up-to-date and, except for Fe, are also used in the
recent ionization equilibrium calculations of \citet{Mazzotta+al:1998}.
We account for direct ionization as well as for excitation-autoionization,
where appropriate.

At each grid point of the MHD model we first evaluate the ionization rate
$C_{\rm{ion}}$ for the respective temperature and obtain the ionization
time by multiplying with the electron density from the MHD model,
\begin{equation}\label{E:ion_time}
\frac{1}{\tau_{\rm{ion}}} ~=~ n_e\,C_{\rm{ion}} ~
\end{equation}
This is done for each ion of the lines listed in \tab{T:lines}.

As we are interested in the effects of the (non-equilibrium) ionization on
the emission lines, we calculate an ionization time for each line by
considering only those grid points of the MHD calculation with
temperatures $\pm0.1$\,dex in $\log{T}$ from the line formation
temperature $T_{\rm{line}}$.
We use values of $T_{\rm{line}}$ as derived in the previous section
(cf.\ column ``MHD'' in \tab{T:lines}).
The range of temperatures was chosen to be $\pm0.1$\,dex, i.e.\ 0.2 in
$\log{T}$ according to the widths of the contribution function (cf.\
FWHM in \tab{T:lines}).

\begin{figure}[t]
\includegraphics[width=\figwcol]{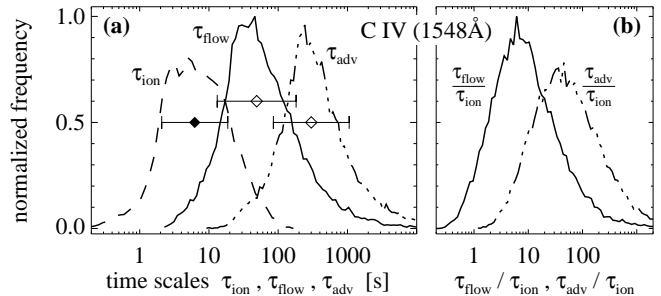}
\caption{%
The left panel (a) shows a histogram of ionization times $\tau_{\rm{ion}}$
for C$^{3+}$ (dashed) as compared to histograms of flow times
$\tau_{\rm{flow}}$ (solid) and advective times $\tau_{\rm{adv}}$
(dashed-dotted) at the temperatures where \ion{C}{4} (1548\,\AA) is
formed.
The diamonds show the median values of the respective times and the bars
indicate the standard deviations.
\newline
The right panel (b) shows the histograms of the ratios of flow to
ionization time, $\tau_{\rm{flow}}/\tau_{\rm{ion}}$ (solid), and advective to
ionization time, $\tau_{\rm{adv}}/\tau_{\rm{ion}}$ (dashed-dotted).
See \sect{S:ionisation}.
\label{F:histo_times}}
\end{figure}

\begin{figure}[t]
\includegraphics[width=\figwcol]{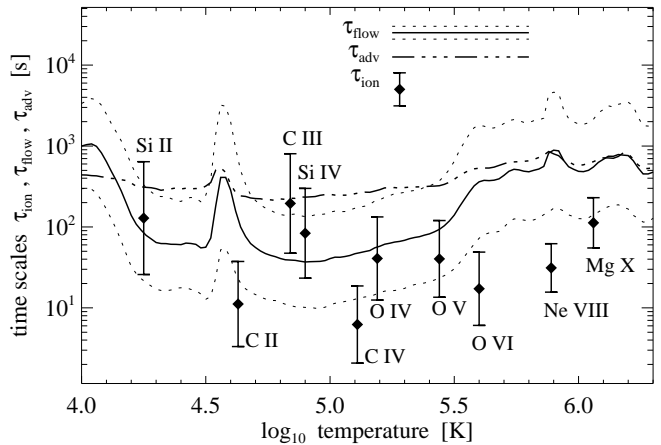}
\caption{%
Comparison of time scales for ionization, flow and advection.
The diamonds show the median values of the ionization time
$\tau_{\rm{ion}}$ of the respective lines (standard deviation shown as
bars, cf.\ \fig{F:histo_times}a).
The solid line shows the median flow time $\tau_{\rm{flow}}$ at the
respective temperature in the MHD model.
The dotted lines indicate the standard deviation of $\tau_{\rm{flow}}$.
In addition the dashed-dotted line shows the median values of the
advective times $\tau_{\rm{adv}}$.
See \sect{S:ionisation}.
\label{F:compare_times}}
\end{figure}

This leads to a distribution of ionization times in the line forming
region of the respective line.
For the case of \ion{C}{4} (1548\,\AA) this histogram is shown as a dashed
line in \fig{F:histo_times}a.
As the ionization time for this line we use the median value of the
distribution (solid diamond).
To characterize the width of the distribution we compute the standard
deviation of the ionization times on a logarithmic scale.
This is shown as a bar in \fig{F:histo_times}a.
The ionization times for the other lines are displayed as just those bars
with diamonds for the median values in \fig{F:compare_times}.

\subsubsection{Dynamic times}                           \label{S:flow_times}

We use two procedures to determine a dynamic time, the first closer linked
to the picture of particles getting ionized while flowing up a temperature
gradient, the second more formally bound to the advective term in the rate
equations.

The flow time $\tau_{\rm{flow}}$ we define as the time needed
for a test particle with velocity $v$ to cross a given temperature
difference $(\Delta\log{T})_{\rm{ion}}$.
Thus within the time $\tau_{\rm{flow}}$ we put the test particle to a
higher temperature and ask if the particle is ionized after this time.
If this is not the case, i.e. if $\tau_{\rm{flow}}>\tau_{\rm{ion}}$,
ionization equilibrium is a good approximation.

We choose a value of $(\Delta\log{T})_{\rm{ion}}=0.1$, as this is about
the half width at half maximum of the contribution function for the
emission lines (FWHM/2 in \tab{T:lines}).
When moving an ion by such a temperature difference it just starts
``seeing'' another ionization equilibrium.

Thus the inverse of the flow time is given by
\begin{equation}\label{E:flow_time}
\frac{1}{\tau_{\rm{flow}}} ~=~
        \frac{{\nabla}(\log{T}) \cdot \bm{\hat{v}}}%
             {(\Delta\log{T})_{\rm{ion}}} ~ |\bm{v}| ~,
\end{equation}
%
%
where the temperature gradient and the velocity vector $\bm{v}$ are
taken from the MHD model.
$\bm{\hat{v}}$ is the unit vector along $\bm{v}$.
The temperature gradient is projected on the flow direction as we are
interested only in changes along the path of the test particle.
Furthermore we consider only locations where the flow is in the direction
of increasing temperature as only this will be affected by the ionization
processes.

Like for the ionization times we now compute the distribution of flow times
in each line formation region.
These histograms can be directly compared to those of the ionization
times, as it is done in \fig{F:histo_times}a for \ion{C}{4}.
This shows that for \ion{C}{4} the flow time is larger than the ionization
time, which is also confirmed by the histogram of ratios of flow to
ionization times (\fig{F:histo_times}b).

In \fig{F:compare_times} the median values of the flow times,
$\tau_{\rm{flow}}$, are shown also for other temperatures in the MHD
model as a solid line.
The dotted lines indicate the standard deviation of $\tau_{\rm{flow}}$.
As expected the flow time is smallest in the middle transition region
below $10^5$\,K as there the temperature gradient is steepest.
$\tau_{\rm{flow}}$ increases by almost two orders of magnitude towards
the corona as well as to the chromosphere.
One should note that the absolute values of the flow times depend on the
choice of $(\Delta\log{T})_{\rm{ion}}$.

\medskip

A more direct way to compute a dynamic time is to evaluate the
left-hand-side of the rate equations \eqn{E:rate_equations}.

If considering only the process of electron collisional ionization for a
single ionization state, the rate equation \eqn{E:rate_equations} reads
\begin{equation}
  \frac{1}{n}~\frac{{\rm{d}}n}{{\rm{d}}t}
+ \frac{1}{n}~\nabla \cdot (n\,\bm{v})  = -n_e~C_{\rm{ion}} ~,
\end{equation}
with $n_i=n$ being the density of the respective ion, $n_j=n_e$ the
electron density and $\gamma_{ij}=n_i\,n_e\,C_{\rm{ion}}$ the ionization
rate.
The right-hand side is the inverse ionization time as used in
\eqn{E:ion_time} and from the left-hand-side we can define an advective
time scale
\begin{equation}\label{E:adv_time}
\frac{1}{\tau_{\rm{adv}}} ~=~ \frac{1}{n}~|\nabla \cdot (n\,\bm{v})| ~,
\end{equation}
The divergence of the particle flux density $\nabla \cdot (n\,\bm{v})$ is
calculated from the MHD model.
Just like for $\tau_{\rm{flow}}$ we have computed the median values of
$\tau_{\rm{adv}}$ and over-plotted them in \fig{F:compare_times} as a
dashed-dotted line.
At all temperatures the advective time scale is at least as large as the
flow time scale defined above, $\tau_{\rm{adv}} > \tau_{\rm{flow}}$.

\subsubsection{Comparing ionization and dynamic times}  \label{S:compare_times}

A comparison of the ionization time $\tau_{\rm{ion}}$ with the dynamic times
$\tau_{\rm{flow}}$ and $\tau_{\rm{adv}}$ shows that in general the
assumption of ionization equilibrium is satisfied in the model under
investigation here.

A more detailed comparison of the distribution of ionization and dynamic
times shows that there are regions where one definitely should not use
ionization equilibrium.
Mostly, like for \ion{C}{4} only a small fraction of the volume has flow
times smaller than ionisation times (cf.\ in \fig{F:histo_times}b only a
minority has values of $\tau_{\rm{flow}}/\tau_{\rm{ion}}{<}1$).
In our study it is only for \ion{C}{3} and \ion{Si}{4} that the assumption
of ionisation equilibrium is violated in the sense that
$\tau_{\rm{flow}}{<}\tau_{\rm{ion}}$, but even then the ionisation time is
still smaller (or comparable) to the the advection time $\tau_{\rm{adv}}$
(cf.\ \fig{F:compare_times}).

From this we can conclude that for the present work and as a first step for
a spectroscopic analysis of complex 3D MHD coronal models we might well
use ionization equilibrium.
Future work, however, should try to include also non-equilibrium effects.

\subsubsection{Ionization equilibrium and flows}    \label{S:ion_flows}

It is often argued that departures from the ionization equilibrium become
of vital importance as soon as the velocities across a temperature gradient
become ``large enough''.
As shown in the preceding subsection in our computation the ionization
times are (generally) still shorter than the time scales of the flows,
and hence the assumption of ionization equilibrium is not too bad.

This is in contrast to earlier considerations, e.g. of
\cite{Joselyn+al:1979}, who argued that velocities of the order of
10\,km/s will certainly lead to a violation of ionization equilibrium.
They used the temperature gradients from \emph{static} models
\citep[e.g.][]{Gabriel:1976} to calculate time scales of the plasma
flowing across these temperature gradients.
However, such an analysis does not account for the back reaction of the
flow on the temperature gradient, and thus a too small time scale is
derived.

\begin{figure}[t]
\includegraphics[width=\figwcol]{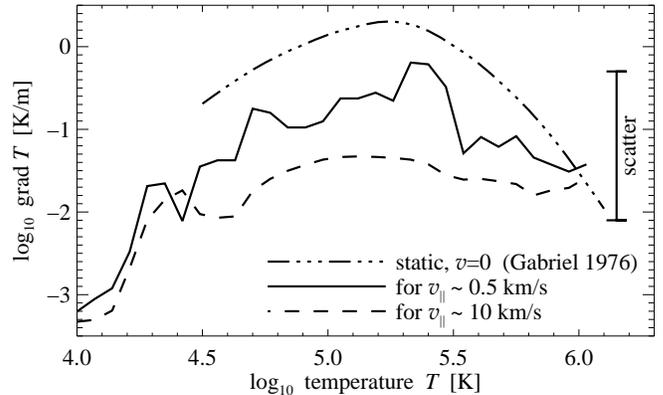}
\caption{%
Average temperature gradient along the streamlines in the MHD model
underlying the present analysis as a function of temperature for regions
with flow speeds of about 10\, km/s (dashed) and 0.5 km/s (solid).
For comparison the gradient in the \emph{static} model of
\cite{Gabriel:1976} is shown as a dot-dashed line.
The bar represents the scatter of gradients in each temperature bin.
See \sect{S:ion_flows}.
\label{F:gradT}}
\end{figure}

To illustrate this we plot the average temperature gradient (along
streamlines)  as a function of temperature within the box of the MHD model,
once for only those regions with velocities of about 10\,km/s (dashed) and
about 0.5\,km/s (solid) in \fig{F:gradT}.
It is clearly evident that the temperature gradient is smaller in regions
with higher velocities.
For comparison we plot the temperature gradient in a static coronal
(funnel) model \citep[dot-dashed in \fig{F:gradT};][]{Gabriel:1976}.
In the case of small velocities the temperature gradient in the MHD model
underlying the present analysis is almost as steep as in a static model
(factor $\sim$2 in the middle transition region).
As there is no heating in the transition region of the static model
of \cite{Gabriel:1976}, while in the 3D MHD model with
flux-braiding the heating is strongly concentrated at low heights, one
expects this somewhat shallower gradient in the latter case.
The main point to stress here is that if the flow speed is larger, the
gradient becomes much flatter, by about an order of magnitude, as the flow
is smoothing the temperature gradient (cf.\ solid and dashed line in
\fig{F:gradT}).

From this discussion one might draw the following conclusion.
For a higher velocity the temperature gradient will decrease,
which partly compensates the effect of a faster speed when calculating
the flow time using e.g.\ \eqn{E:flow_time}.
By this the back reaction of the flow on the temperature helps in
simplifying the treatment of ionization to the case of equilibrium.
However, for models with more violent flows, the assumption of ionization
equilibrium certainly will break down.

\subsection{DEM from the MHD calculations}              \label{S:dem_mhd}

To check the validity of the inversion of the differential emission
measure ({\DEM}) discussed in \sect{S:dem} we calculate the {\DEM} directly from
the MHD results, i.e.\ the density.
For a traditional {\DEM} inversion one has to assume a 1D atmosphere with a
monotonic increase in temperature.
Only then the {\DEM} definition in \eqn{E:dem} makes sense.
However, this is not the case in a highly structured atmosphere as under
investigation here.
Thus we start by defining a volume emission measure
\begin{equation}\label{E:EMv}
{\EM}_V \, (T) ~=~ \int_V n_e^2~{\rm{d}}V ~,
\end{equation}
where the integration is over a volume typically contributing to an
emission line formed at temperature $T$.
According to the contribution functions to the line
(\sect{S:lineformation}) the volume covers a temperature interval of
${\pm}0.15$ in ${\log}T$, i.e.\ ${\Delta}{\log}T{=}0.3$.

To move to a height-related expression the volumetric emission measure is
divided by the area $A$ of the  box giving an {\em average} column emission
measure ${\EM}_h = {\EM}_V/A$.
%
%
%
%
Now one can formally substitute the volume integration by a height
integration ${\rm{d}}h={\rm{d}}V/A$ and this by an integration over
temperature,
\begin{equation}\label{E:EMh_dT}
{\EM}_h = \int_h n_e^2~{\rm{d}}h
\quad \Leftrightarrow \quad
{\EM}_h = \int n_e^2~\frac{{\rm{d}}h}{{\rm{d}}T}~{\rm{d}}T ~.
\end{equation}
Using the definition of the {\DEM} \eqn{E:dem} one gets
\begin{equation}\label{E:EMh_DEM}
%
{\EM}_h = \int {\DEM} ~{\rm{d}}T ~.
\end{equation}
Thus one can approximate the {\DEM} by
\begin{equation}\label{E:DEM_MHD}
{\DEM}_{\rm{MHD}} ~=~ \frac{{\EM}_h}{\Delta T} ~=~ \frac{{\EM}_V}{A\,\Delta T} ~,
\end{equation}
where ${\Delta}T$ is the temperature range corresponding to
${\Delta}{\log}T{=}0.3$, i.e. ${\Delta}T$ increases with temperature.

Using \eqn{E:EMv} this allows the calculation of the
${\DEM}$ directly from the MHD model, which can be compared to the ${\DEM}$ as
resulting from the inversion of line intensities as outlined in
\sect{S:dem}.

In \fig{F:dem} we plot the {\DEM} derived directly from the MHD model as
following \eqn{E:DEM_MHD} as a dot-dashed line along with the {\DEM} from the
inversion using the line emissivities (solid) as discussed in
\sect{S:dem}.
We see that they compare relatively well, even though the difference at
high temperatures is noticeable.
As this discrepancy is probably (at least partly) due to the highly
structured nature of the atmosphere, a further discussion is shifted to
\sect{S:dem_results}.

\begin{figure*}[t]
\centerline{\includegraphics[width=\figwtxt]{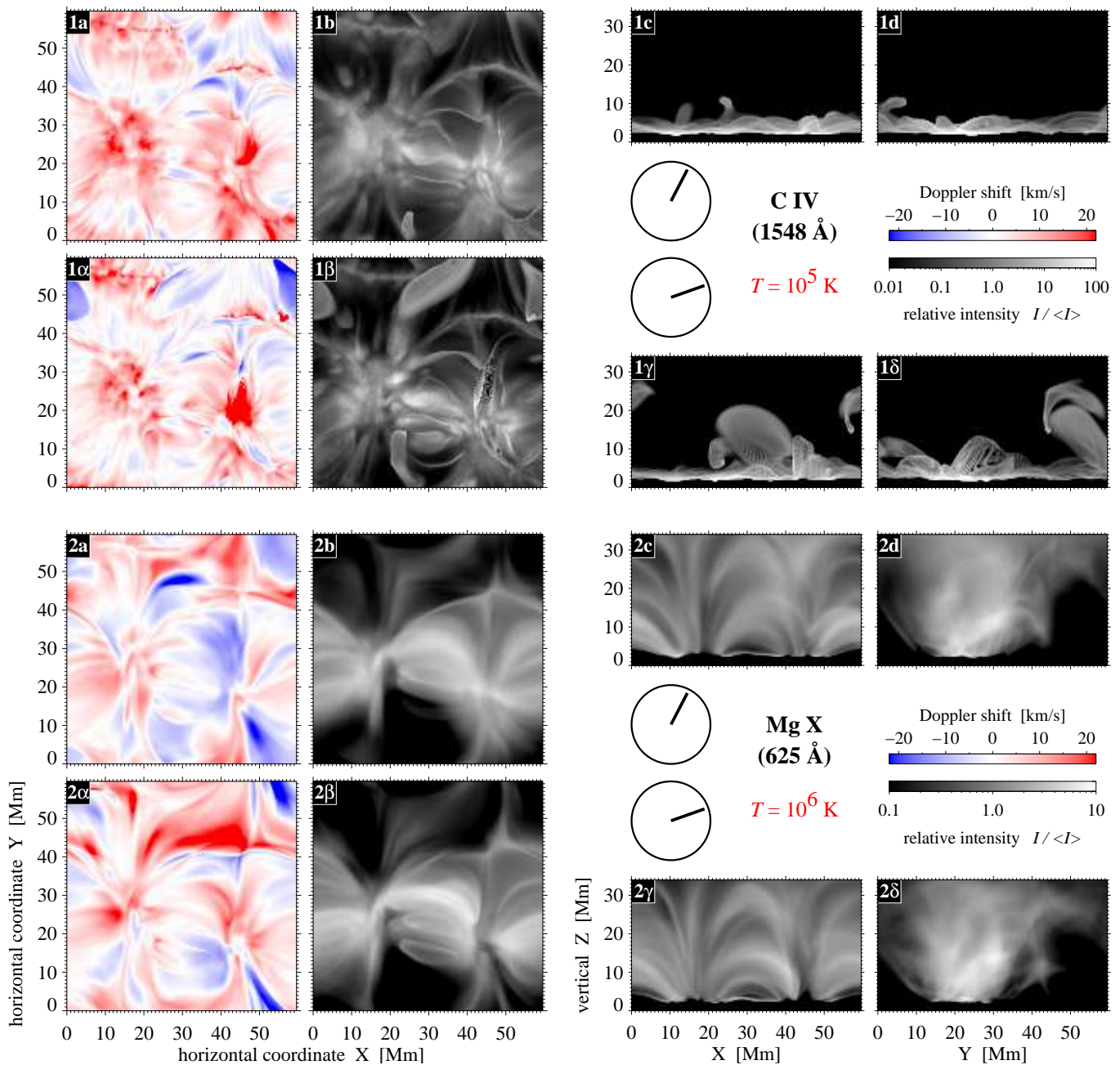}}
\caption{%
Spatial maps in Doppler shift and intensity in the lines of \ion{C}{4}
(1548\,\AA) and \ion{Mg}{10} (625\,\AA) formed at about $10^5$ and
$10^6$\,K, respectively, for two different time steps of the MHD model
about 7 minutes apart (at $t{=}4$\, min and 11\,min).
\newline
The two top panels labeled with 1 show \ion{C}{4}, the two bottom panels
labeled with 2 show \ion{Mg}{10}.
The maps of the earlier time step are labeled with Latin letters (1st and
3rd row), those for the time step 7 minutes later with Greek letters (2nd
and 4th row).
\newline
The panels (a/$\alpha$) show the Doppler shift of the synthesized spectra
as seen from straight above, the panels (b/$\beta$) show the same for the
intensity of the respective line.
This corresponds to the appearance near disk center.
Panels (c/$\gamma$) and (d/$\delta$) show side views of the computational
box along the x and y axis in line intensity, which resembles the
appearance at the limb.
\newline
The intensities $I$ are scaled with respect to the average (median) intensity
$\langle{I}\rangle$ of the respective map.
\label{F:image}}
\end{figure*}

\section{Results}               \label{S:analysis}

Here we will only give some examples of the results that may be achieved
using the spectroscopic analysis described in this paper.
More specific problems, such as time variability, evolution or spatial
structure will be addressed in future investigations and will provide new
tests for the model of coronal heating through flux braiding.

\subsection{Morphology: intensity and Doppler shift maps} \label{S:maps}

By investigating maps in intensity and Doppler shift in various lines
when integrating along a line-of-sight through the box, one can now study
the morphology of the transition region and corona.
Such maps are provided in \fig{F:image} for \ion{C}{4} formed in the low
transition region at ${\sim}10^5$\,K (top rows labeled with 1) and
for \ion{Mg}{10} formed in the low corona at ${\sim}10^6$\,K (bottom rows
labeled with 2).
There the respective leftmost panels (a/$\alpha$) show the Doppler maps
when looking from straight above and the panels b/$\beta$ show the same in
intensity --- this would correspond to an observation at disk center.
The right panels (c/$\gamma$, d/$\delta$) show intensity maps when looking
from the sides at the box (displayed with correct horizontal/vertical
aspect ratio) --- this corresponds to the appearance at the limb.
For each line the snapshots at two times are shown at some 4 and
11\,minutes into the simulation (labeled by Latin and Greek letters,
respectively).

In an earlier publication we reported shortly on the more structured
appearance of the transition region being due to heat conduction and the
numerous structures in coronal Doppler shifts \citep{Peter+al:2004}.

When comparing the intensity maps of the two time steps only seven minutes
apart, it is striking how much the transition region has changed in this
time, while the corona still looks pretty much the same in intensity.
This is of special interest, if one recalls that the driver of the corona,
the photospheric granular motion, acts on the time scale of some 5 to 10
minutes.
Thus the transition region directly responds to the changes in the
photospheric magnetic structure, while the corona is reacting much slower.
This difference is mainly due to the large difference in cooling time
scale and the very efficient re-distribution of energy trough heat
conduction at high temperatures.

Not surprisingly also the Doppler map of the transition region changes
significantly during the only seven minutes on small scales.
Despite the smooth, slowly-evolving appearance in intensity, the
corona reveals its dynamic structure in the Doppler map.
While the heat conduction quickly distributes the deposited energy, still
the dynamic response of the plasma to the heating process is present,
showing up as flows causing the Doppler shifts.
And just like in the transition region, the coronal Doppler shifts show
large variations on small spatial scales within the time scale of the
photospheric driver.

This high variability of the coronal Doppler maps shows the need to study
coronal variations on short time scales not only using intensity maps like
provided by EIT/SOHO or TRACE.
If one could obtain Doppler maps with high temporal resolution, one would
have access to the dynamic response of the corona to the heating process,
which is not available through intensity maps alone.
Unfortunately current slit spectrographs like SUMER or CDS on SOHO do not
provide maps of sufficient size in a reasonable time cadence.

\begin{figure*}[t]
\centerline{\includegraphics[width=\figwtxt]{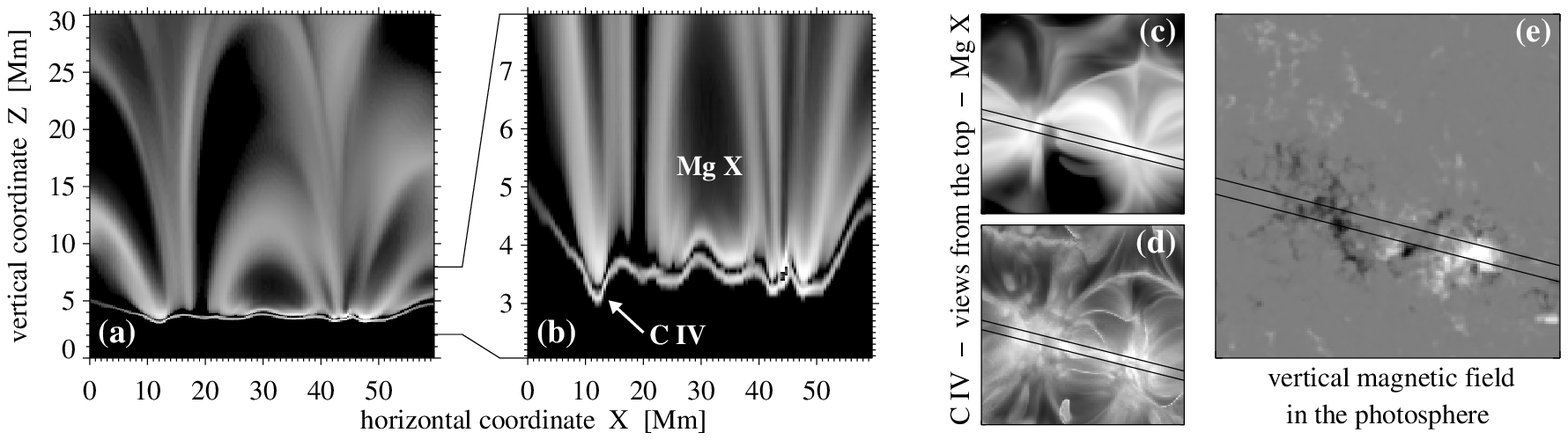}}
\caption{%
Composite view of a thin vertical slice of the computational box in
\ion{C}{4} (${\sim}10^5$\,K) and \ion{Mg}{10} (${\sim}10^6$\,K) at time
$t{=}4$\,min.
The left panel (a) shows the vertical slice as defined by the black lines
in the top views (c,d) when integrating along the Y
direction; actually panels (c,d) are the same as panels (1b,2b) of
\fig{F:image}.~
The location of the slice with respect to the photospheric magnetic field
map is shown in panel (e).
The middle panel (b) shows part of panel (a) stretching the vertical axis
to highlight the structure in the low parts of the atmosphere.
The emission as seen in \ion{C}{4} and \ion{Mg}{10} comes from disjunct
areas of the vertical slice.
The lower ``line'' shows the emission from \ion{C}{4}, while the upper
more diffuse emission stems completely from \ion{Mg}{10}.
%
%
See \sect{S:maps:warp}.
\label{F:cut}}
\end{figure*}

\subsection{Vertical structure: roiling and loops}  \label{S:maps:warp}

To show the further potential of the forward modeling technique presented
here, we investigate a vertical slice of the computational box cutting
the two main polarities forming the active region.
In \fig{F:cut} c, d, and e we show the location (between the lines) of the
vertical slice in the horizontal plane for the intensity maps in \ion{C}{4}
and \ion{Mg}{10} as well as the photospheric magnetogram.
The left panel (\fig{F:cut}a) shows the vertical slice, the emission from
\ion{C}{4} and \ion{Mg}{10} displayed on top of each other.
Please note that in contrast to \fig{F:image} here the aspect ratio is not
unity, but about a factor of two.

The emission from these two lines comes from practically disjunct regions.
The coronal \ion{Mg}{10} emission stems from the diffuse loop-dominated
part in the upper part of the image, while the transition region
\ion{C}{4} emission originates from a very thin layer below the corona.
Panel b of \fig{F:cut} shows the area from ${\sim}2$--$8$\,Mm
height vertically stretched to show the transition region more clearly.
At least in this region the transition region emission comes from an area
below the coronal source region, i.e. from the footpoint regions of the
coronal magnetic structures.

\fig{F:cut}b also reveals that the transition region is roiling field,
while being very thin in the vertical direction (some 100\,km).
This is to be expected, as the heating is highly variable in space and time
and thus the location of the transition region is to be expected to move
up and down to adjust the coronal pressure in accordance with the heating
rate.
This is well known from 1D loop models \citep[e.g.][]{Hansteen:1993}
and also commented on by \cite{Gudiksen+Nordlund:2005a}.
%
Despite being very thin, the roiling of the transition region leads to a
relatively thick appearance when integrating along a horizontal direction.
In \fig{F:cut}b the roiling of the transition region has an amplitude in
height of about 1\,Mm or more, while at other places and times it can be
even more.
Thus the roiling leads to a transition region that would appear several Mm
thick when observed at the limb, as nicely illustrated by the side views
in \fig{F:image}, panels 1c and 1d.
This corresponds well will observations of limb intensity profiles at the
limb \citep[e.g.][]{Mariska+al:1978}.

\subsection{Average line shifts} \label{S:line_shifts}

\begin{figure}[t]
\centerline{%
    \includegraphics[width=\figwcol]{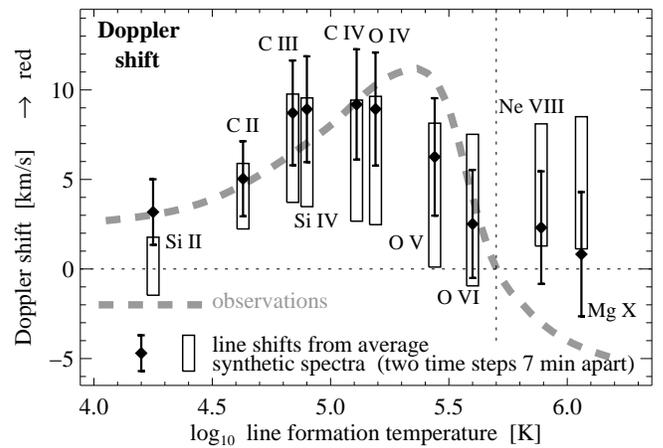}}
\caption{%
Comparison of synthesized and observed Doppler shifts for two time steps
about 7 minutes apart (same times as in \fig{F:image}; $t{=}4,11$\,min).
The diamonds show the Doppler shifts of the average spectra and the bars
indicate the standard deviation of the Doppler shifts of a spatial map as
seen looking from straight above on the computational box for the earlier
time step.
The rectangles show the same but for 7 minutes later.
The change in this short time is noticeable.
The lines are plotted at the formation temperature as following from the
contribution to the emission from the present work (cf. \tab{T:lines}).
The thick dashed line shows the trend as found in observations.
See \sect{S:line_shifts}.
\label{F:line_shifts}}
\end{figure}

The time-average of the line shifts derived from the synthetic corona
presented here were discussed previously \citep{Peter+al:2004}.
In \fig{F:line_shifts} we plot the average line shifts as a function of
line formation temperature for the two individual time steps, seven minutes
apart, which have been displayed also in the spatial maps in \fig{F:image}.
The Doppler shifts are calculated for a vertical line-of-sight, i.e.\ when
looking at the computational box from straight above.
The height of the bars and rectangles represents the spatial scatter for
each time step (1/2 of the standard deviation).
The trend found in observations is plotted as a thick dashed line
\citep[compiled from][]{Brekke+al:1997,Chae+al:1998,Peter+Judge:1999}.

It is obvious that the average Doppler shifts change quite
dramatically, and the overall shape varies from being very close to the
observations (bars) to a more flat profile with comparable Doppler shifts
at all temperatures (rectangles) --- and this happening in only seven
minutes.
The time variation of the Doppler shift for some 20 minutes is shown in
\fig{F:temporal}b for a number of lines.
Here we see that for all lines, also for the coronal \ion{Mg}{10} line,
the variation over these 20 minutes is comparable or larger than the
spatial scatter in the data.
This again emphasizes the potential provided by observations of Doppler
shifts throughout the outer solar atmosphere, as pointed out at the end of
\sect{S:maps}.

In the present model the persistent redshifts in the transition region,
puzzling theorists since their discovery by \cite{Doschek+al:1976},
are caused by the flows induced by the heating through braiding of magnetic
flux.
Being free of spurious assumptions, this is the first time one finds an
explanation of the Doppler shifts providing a qualitative and quantitative
representation of the redshifts in the  middle transition region.
The blue shifts towards higher temperatures are not matched yet, but we see
a clear indication that above $\log{T}{\approx}5.2$ the redshifts are
decreasing, and at some periods of time in the model run, we are quite close to the
observations (cf.\ bars in \fig{F:line_shifts}).
Further investigations, especially with an improved upper boundary condition
or a larger box, will have to address this discrepancy.

\subsection{Average non-thermal line widths} \label{S:non_thermal}

\begin{figure}[t]
\centerline{%
    \includegraphics[width=\figwcol]{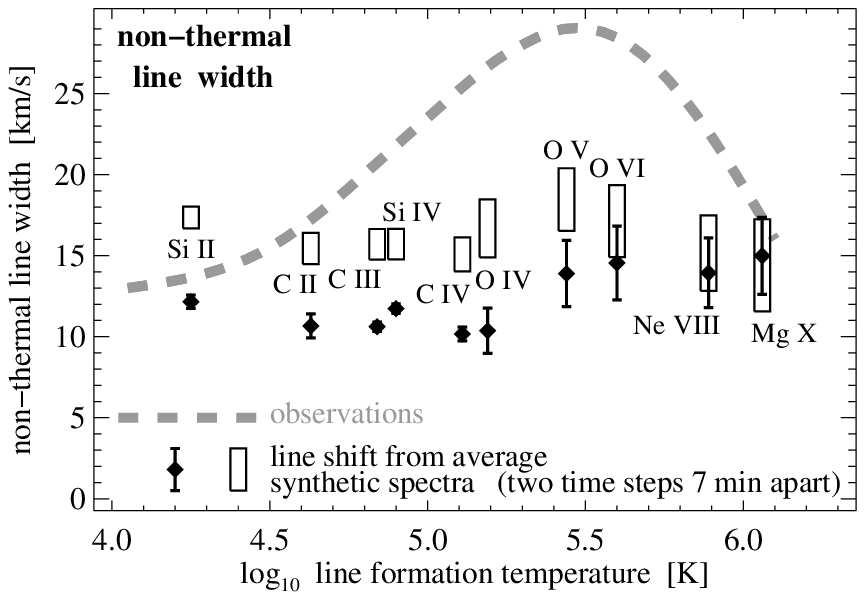}}
\caption{%
Comparison of synthesized and observed non-thermal widths.
Similar as \fig{F:line_shifts} but now for non-thermal widths.
See \sect{S:non_thermal}.
\label{F:non_thermal}}
\end{figure}

Another important test for any coronal model is provided by a comparison
of the non-thermal line widths.
These contain information on the non-thermal unresolved motions,
introduced by the limited resolution  in space and time of any
instrument.
Averaging in space and time will mix spectral information with different
velocities and will lead to a broadening of the line.
It is important to note that also the integration along the line-of-sight
contributes to the non-thermal broadening, and there is (almost) nothing we
can do about this from an observers point of view.

In \fig{F:non_thermal} we plot the non-thermal line width of the average
spectrum for the two time steps seven minutes apart already displayed in
\fig{F:image} and \ref{F:line_shifts} (again when looking at the
computational box from straight above).
The thick dashed line shows the trend in observations compiled from
\citet{Chae+al:1998:width} and \citet{Peter:2001:sec}.

In the corona and the low transition region the non-thermal width from the
synthesized spectra match roughly the observed values, but in the middle
transition region the synthetic spectra show significantly smaller widths
than the observations.
There is (at least) one possible explanation for this.
The high non-thermal broadening observed on the Sun might be due to
small scale velocities connected with the heating process itself, i.e. the
nano-flares induced by the flux-braiding.
And this effect is to be expected to have the largest effect in the middle
transition region, where the time scales are shortest.
Because of their tiny spatial scale, these nano-flares cannot be
modeled in the MHD simulation, which the spectral synthesis is based upon.
Future models with increased spatial resolution will have to show if this
mismatch with the observations can be resolved.

\subsection{What do we learn from the differential emission measure?}
                                                        \label{S:dem_results}

\begin{figure}[t]
\centerline{\includegraphics[width=\figwcol]{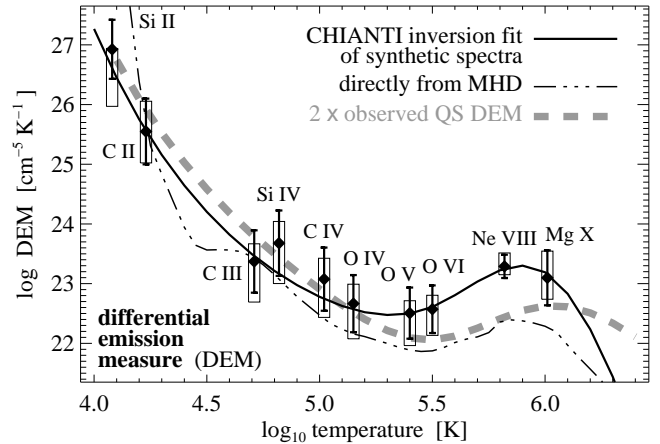}}
\caption{%
Comparison of the differential emission measure ({\DEM}) as derived from the
synthesized and observed spectra.
The solid line shows the fit from the {\DEM} inversion based on the
synthesized lines displayed as bars (\sect{S:dem}; for time $t{=}4$\,min).
The lines are displayed here at the formation temperature as following
from the {\DEM} inversion (cf. \tab{T:lines}). This differs slightly from the
temperatures used in Figs.\ \ref{F:line_shifts} and \ref{F:non_thermal}.
The rectangles show the situation about 7 minutes later at $t{=}11$\,min,
the changes being very small.
The thick dashed line is based on a {\DEM} inversion using observed quiet Sun
disk center line radiances observed with SUMER, scaled by a factor of two.
The dot-dashed line displays the {\DEM} curve as following directly from
the MHD results (\sect{S:dem_mhd}).
See \sect{S:dem_results}.
\label{F:dem}}
\end{figure}

As outlined in \sect{S:dem} and \ref{S:dem_mhd}, we perform an
differential emission measure ({\DEM}) analysis using the spectral lines
listed in \tab{T:lines} using the atomic data package CHIANTI.
The resulting {\DEM} curve for one single time step (at 4\,minutes into the
simulation) is plotted in \fig{F:dem} as a solid line.
The diamonds show the {\DEM} at the line formation temperature of the
respective line, multiplied by the ratio of the observed emissivity to the
one predicted by the {\DEM} fit.
Thus the displacement of the diamonds from the solid line represent the
error of the {\DEM} fit.
The height of the bars represent the scatter of the emissivities.

As we discussed previously \citep{Peter+al:2004} this is the first model
to reproduce the overall shape of the {\DEM} curve quantitatively and
qualitatively, especially the increase of the {\DEM} towards lower
temperatures below $\log{}T{=}5.3$.
For comparison we plotted an {\DEM} inversion based on radiance data at disk
center taken from \citet{Wilhelm+al:1998}, for \ion{Mg}{10} and
\ion{Si}{2} we re-evaluated some SUMER disk center spectra.
The agreement is remarkable, except for the highest temperatures.
This discrepancy is partly due to a lack of constraining
lines at higher temperatures, as the maximum temperature in the MHD
simulation is about $\log{T}{\approx}6.3$.
Thus the {\DEM} inversion is not very well defined at high temperatures.
This becomes especially important when discussing the temporal variability
of the {\DEM} below.

The {\DEM} inversion based on the synthesized spectra roughly agrees with the
{\DEM} derived directly from the MHD models (cf.\ \sect{S:dem_mhd}).
While for the {\DEM} inversion one implicitly assumes a simple 1D stratified
static atmosphere (cf.\ \sect{S:dem}) this is certainly not the case in
the computational box.
In this light it is even surprising that the {\DEM} derived directly from the
MHD model (\fig{F:dem}, dot-dashed line) roughly compares to the
inversion, especially below $\log{T}{=}5.0$.

\begin{figure}[t]
\centerline{\includegraphics[width=\figwcol]{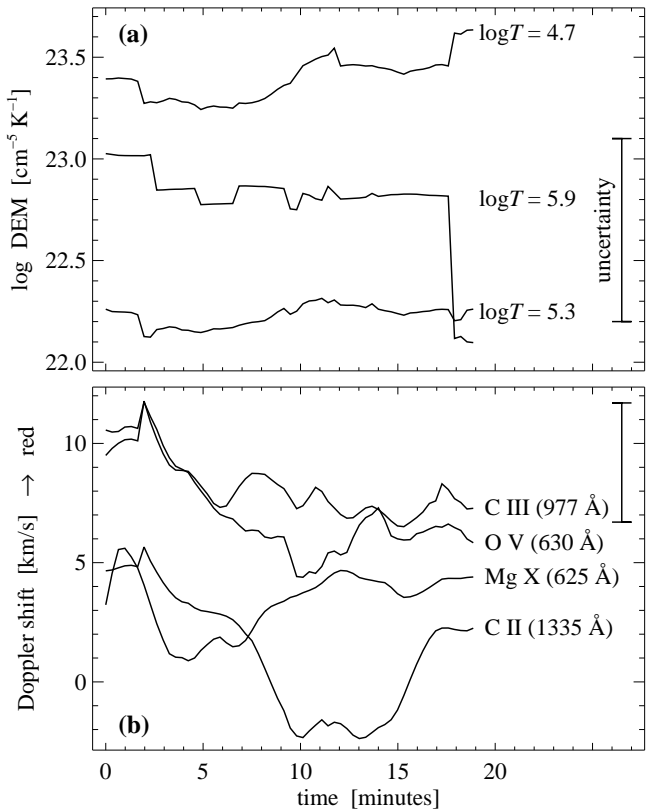}}
\caption{%
Temporal evolution of the differential emission measure ({\DEM}) and the
average Doppler shift.
The top panel (a) shows the evolution for the {\DEM} at three different
temperatures as following from {\DEM} inversion fits like the one shown in
\fig{F:dem}.
The bottom panel (b) shows the variation of the Doppler shifts for
emission lines spanning temperatures from $\log{T}{\sim}4.6$ to $6.0$.
Please note that the variation in {\DEM} in much smaller than the
uncertainties (represented by the bars in both panels) and is thus not
significant.
In contrast, the variation in Doppler shift is significant.
The Figs.\,\ref{F:image} -- \ref{F:dem} show snapshots at $t{=}4$\,min and
11\,min.
See \sect{S:dem_results}.
\label{F:temporal}}
\end{figure}

In \fig{F:dem} the rectangles represent the situation concerning the {\DEM}
inversion 7\,minutes after the time step discussed so far (represented by
the bars).
The change is very small.
This is emphasized by the temporal evolution of the {\DEM} inversion fit at
temperatures representing the low and middle transition region as well as
the low corona (\fig{F:temporal}a).
It is clearly evident that the changes in the {\DEM} fit are only very small.
At $\log{T}{=}5.9$ the fit is ``jumping'', as the {\DEM} inversion is not
well constrained at high temperatures (see above).
Thus also the jump of the fit for $\log{T}{=}5.9$ at time
${\sim}18$\,minutes is an artefact and not real.

The important result here is that the {\DEM} hardly changes with time
(over the 20 minutes of the simulation), but within the uncertainties
stays constant.
This is in strong contrast to the Doppler shifts, which change significantly
at all temperatures.

The {\DEM} curve thus provides an important test for a coronal heating
model, in particular in the sense that it is not easy to get the overall
shape right.
However, once one gets the overall shape right, it seems that the {\DEM} stays
fixed, and therefore provides only limited potential for further
diagnostics of the structure and dynamics of the coronal heating process.

\section{Conclusions}           \label{S:conclusions}

We have presented a procedure how to synthesize EUV spectra from a complex
3D MHD model for a stellar corona, in which the heating is due to flux
braiding through photospheric footpoint motions.
An investigation of the validity of the ionization equilibrium showed that
in the present case this simplification is justified.
Because the heating is concentrated very low in the atmosphere, the
temperature gradients are much shallower than in traditional static
models, and thus the typical dynamic (or flow) times are larger than the
ionization times.
However, for models with more violent flows one will have to properly
solve the ionization rate equations.
We found that due to the strong heating at the footpoints of the
corona and the resulting strong increase of density towards the
chromosphere, cool lines like from \ion{Si}{2} tend to be formed well
below their ionization equilibrium temperature.
Thus the cool lines are of limited diagnostic value for the transition
region.

Together with the spectral synthesis the MHD simulation provides a forward
model, which allows a direct comparison to observations, and by this is a
powerful tool to investigate the corona.
In this paper we describe some of the potentials of this method.

The spatial maps of intensity show a much higher spatial and temporal
variability in the transition region than in the corona.
In the transition region the temporal variability occurs on a time scale
compatible with the photospheric driver of the coronal heating.
In the Doppler shift maps, however, the spatial variability is also very
large in the corona, much stronger than the variability in intensity.
This shows the need for further instrumentation to get access to coronal
Doppler maps with sufficient temporal resolution (below 5 minutes).

Likewise the time variation of the Doppler shifts is significant, while
the differential emission measure ({\DEM}), derived from the intensities, shows
only a very small variation.
While this model is the first to reproduce qualitatively and
quantitatively the form of the {\DEM} variation with temperature, it also
shows the limitations of the {\DEM} analysis in a highly dynamic atmosphere.
To understand the dynamics of the corona it is vital to use the Doppler
shifts as a crucial test for the model, and not only the line intensity or
emission  measures.

This study shows the pivotal importance of forward modeling for our
understanding of stellar coronae, as it provides numerous controllable ways
to compare model results and observations.


\begin{acknowledgements}
The work of {\AA}N is supported by grant number xx-xxxx-xx from the
Danish Natural Science Research Council (FNU).  Computing time for
the coronal modeling was provided by the Danish Scientific Computing
Center (DCSC) and the Swedish National Allocations Comittee (SNAC).
Sincere thanks are due to the ongoing efforts of the CHIANTI team to
provide the community with an up-to-date atomic data package.
\end{acknowledgements}

%
%


\end{document}